\documentclass[fleqn,twoside]{article}
\usepackage{espcrc2}
\usepackage{epsfig}
\usepackage{amsmath}
\usepackage{amssymb}

\newcommand{\la}[1]{\label{#1}}
\newcommand{\be}{\begin{equation}}
\newcommand{\ee}{\end{equation}}
\newcommand{\ba}{\begin{eqnarray}}
\newcommand{\ea}{\end{eqnarray}}
\newcommand{\nn}{\nonumber \\}

\newcommand{\eq}{Eq.~}
\newcommand{\se}{Sec.~}
\newcommand{\eqs}{Eqs.~}
\newcommand{\nr}[1]{(\ref{#1})}
\renewcommand{\(}{\left(}
\renewcommand{\)}{\right)}
\newcommand{\lk}{\left[}
\newcommand{\rk}{\right]}

\newcommand{\e}{\epsilon}
\newcommand{\sumint}[1]{\underset{#1}{\scalebox{1}{$\sum$}\hspace{-4.2mm}\int\hspace{1.2mm}}}
\newcommand{\sumintp}[1]{\underset{#1}{\scalebox{1}{$\sum^\prime$}\hspace{-5.2mm}\int\hspace{2.2mm}}}
\newcommand{\vp}{\vec{p}}
\newcommand{\vq}{\vec{q}}
\newcommand{\vr}{\vec{r}}
\newcommand{\intt}[1]{\int_{#1}}

\title{\vspace*{-5mm}\mbox{}\hfill{\small BI-TP 2010/20}\\
Open problems in hot QCD}

\author{Jan M\"oller\address[1]{Faculty of Physics, 
University of Bielefeld, D-33501 Bielefeld, Germany}, 
York Schr\"oder\addressmark[1]\thanks{Talk presented 
at Loops and Legs in Quantum Field Theory 2010,
W\"orlitz, Germany, April 25-30, 2010}}

\begin{document}

\begin{abstract}
We try to give a comprehensive review of the main methods
used in modern multi-loop calculations in finite-temperature
field theory. 
While going through explicit examples, we point out similarities
and differences with respect to the zero-temperature case,
utilizing common techniques in a transparent way 
whenever possible.
\vspace{1pc}
\end{abstract}

\maketitle

\section{Introduction}
\la{se:1}

Perturbative computations in finite-temperature QCD are presently 
being pushed to the 4-loop level, resulting in a large number of
sum-integrals over Feynman propagators to be evaluated.
While the reduction problem can in principle be tackled by 
integration by parts (IBP) methods, 
for which -- due to their prevalence in zero-temperature
calculations -- sophisticated algorithms and public computer programs
are available by now, 
the problem of evaluating the resulting set of master sum-integrals
still presents a formidable challenge.
Concerning the reduction step, note, however, 
that at finite temperature (T) the number of master 
integrals at a given loop order is in principle unbounded, which
can be seen already from the infinite number of massless 
1-loop tadpoles \cite{ll08} (see also \eq\nr{I} below),
a situation completely different from that at $T=0$, where
finiteness can be proven rigorously \cite{finiteMas}.

It is fair to say that the tools needed for a systematic evaluation of 
multi-loop sum-integrals are by far not as evolved as those at zero $T$,
where a number of powerful analytic and numerical methods have been
developed and made available, 
such as Mellin transforms, harmonic sums, difference equations 
or sector decomposition, to name a few.

In contrast, the few sum-integrals that have been computed beyond the 
2-loop level have been solved on a case-by-case basis
(see e.g. \cite{AZ,phi4,AndKyll} and references therein), 
mostly by carefully studying the integral at hand, disentangling
(sub-) divergences by suitably tailored subtractions, and using
mixed numerical and analytic methods to obtain the finite terms.

So to make progress with perturbative finite-temperature 
field theory, it would be most welcome to utilize more
zero-temperature machinery than just the IBP relations.
In this note, as a first small step towards this goal, we intend
to display the issues involved in evaluating a typical nontrivial 
sum-integral in a somewhat modern language, which allows to 
pinpoint parallels as well as key differences to the zero-$T$ case.
Working on a specific 3-loop example, we will re-derive one known
sum-integral ($S_1$ from \cite{phi4}, contributing to the 4-loop
pressure of scalar theory) and, essentially by changing
an index $N$ in the computation, generalize it to a new result,
which will contribute to the matching coefficient $g_E^2$
in the dimensional reduction framework of hot QCD \cite{janDipl}.

After introducing some basic notation and defining a concrete
one-parameter sum-integral that shall serve as the main vehicle
to display the various techniques, we will exhibit the main tools
and ideas needed to systematically dissect the integral into divergent 
(but analytically tractable) and finite (but more difficult) 
pieces in \se\ref{se:3}.
The following two sections deal with those (more difficult)
pieces, deriving simple one-dimensional integral representations
which are then evaluated numerically.
While \se\ref{se:6} is somewhat outside the main flow of the paper, and 
serves to make available a number of useful formulae, 
\se\ref{se:7} contains the main new result.

\section{Notation and preliminary remarks}
\la{se:2}

Perturbative calculations in field theories at non-zero 
temperature (T) can be organized in large parts in exact
analogy with zero-$T$ ones, in particular for situations where
the system is in thermal equilibrium. 
Key differences are the additional scale $T$ involved,
and the manifest breaking of 4-dimensional Lorentz symmetry,
both effects being induced by the presence of a heat bath
to which the system under study is coupled.
As a consequence, the temporal direction is compactified on a 
circle, leading to discretized Fourier modes, which have
to be summed over. Working in dimensional regularization, 
this amounts to changing the familiar integral measure as
\be
\int\frac{d^{4-2\e}q}{(2\pi)^{4-2\e}} \rightarrow
T\sum_{q_0} \int\frac{d^{3-2\e}\vq}{(2\pi)^{3-2\e}}
\nonumber
\ee
while, using Euclidean notation, 
four-momenta are written as $Q=(q_0,\vq)$ with 
$Q^2=q_0^2+\vq^2$ 
where $q_0=2\pi T n$ with $n\in\mathbb{Z}$ the summation index
for the bosonic case, to which we will stick throughout this note.
We will write $d$\/-dimensional results using $d=3-2\e$,
and often use
\be
\intt{\vq} \equiv \int\frac{d^d\vq}{(2\pi)^d} \;.
\nonumber
\ee

As already mentioned in the introduction,
the sum presents a major complication when compared to similar 
integrals at zero $T$. Indeed, the 4d integrals are
contained in the sum-integrals, as can be easily seen by expressing
the sum as a contour integral
\ba
T \sum_{n=-\infty}^{\infty} F(2\pi T n) = 
\int_{-\infty}^\infty \frac{dz}{2\pi}\,F(z) 
+\nn
+\int_{-\infty-i0}^{\infty-i0}\frac{dz}{2\pi}\,
\frac{F(z)+F(-z)}{e^{iz/T}-1}
\nonumber\,,
\ea
which holds for analytic functions $F(z)$ which have 
no poles on the real axis.
The first term, being independent of $T$, 
therefore contains the leading UV behavior of the respective integral.
We will use this fact below in \eqs\nr{piDec} and \nr{piTildeDec}.

Let us now concentrate on the specific example of 
massless 3-loop basketball-type sum-integrals
\be \la{BN1}
B_N \equiv \sumint{PQR} 
\frac1{[Q^2]^N\,(P-Q)^2\,R^2\,(P-R)^2} \;,
\ee
of which the special cases $N\in\{1,2,3\}$ occur as master integrals
in perturbative corrections
e.g. to the 3-loop pressure of hot QCD \cite{AZ}, 
to the 4-loop pressure of scalar theories \cite{phi4}
and to 3-loop matching coefficients \cite{janDipl}, 
respectively.

Guided by the idea that this class of integrals, since they
originate from diagrams with two vertices, require only one integration
in coordinate space (as opposed to three), 
it seems desirable to perform the calculation in $x$\/-space 
whenever possible.
However, divergences (in $\e$ as $d\rightarrow 3$) 
obstruct this simple idea.
The reason is that while the Fourier-transformed propagator
(cf. \eq\nr{ft1} ff) has a simple analytic form in $d=3$,
it is a messy object for general $d$.
Hence, care has to be taken to perform suitable subtractions
for the integral, and only transform to coordinate space in 
(IR- and UV-) finite integrals whose values are then needed 
at $\e=0$ only.
To isolate these finite parts requires a series of rearrangements,
which we will now systematically construct,
using methods pioneered in \cite{AZ}.

\section{Setup: subtractions in $d$ dimensions}
\la{se:3}

We start by separating the dominating large-$P$
behavior of the massless 1-loop propagator
\be \la{piDec}
\sumint{R} \frac1{R^2(P-R)^2} =
\frac{\beta}{[P^2]^\e}
+\frac{2I_1}{P^2}
+\Delta\Pi(P)\,,
\ee
where the leading term $\beta\equiv G(1,1,d+1)$ 
as given in \se\ref{se:6}
is simply a 4d massless 1-loop bubble,
the second term contains the massless 1-loop tadpole 
at finite $T$ and carries a factor of two coming from
two ways of routing the large external momentum $P$
through the propagators,
such that the UV-subtracted remainder $\Delta\Pi(P)\propto 1/[P^2]^2$
as $|p_0|,|\vp|\gg T$.

According to \eq\nr{piDec}, the sum-integral \eq\nr{BN1} 
decomposes as
\ba \la{BN2}
B_N &=& 
\beta\,\sumint{PQ} \frac1{[Q^2]^N[P^2]^\e(P-Q)^2}
\nn&+&
2I_1\,\sumint{PQ}\frac1{[Q^2]^NP^2(P-Q)^2}
\nn&+&
\sumint{PQ}\frac{\Delta\Pi(P)}{[Q^2]^N(P-Q)^2} \;.
\ea

In order to split off potential IR divergences coming
from $1/[Q^2]^N=1/[\vq^2+q_0^2]^N$ when $q_0=0$
(its {\em zero-mode}), 
we multiply each of the three terms by the identity
$(\delta_{q_0}+(1-\delta_{q_0}))$. 
To complete the IR-subtraction for the third term,
we multiply $\delta_{q_0}$ by the
identity $(\delta_{p_0}+(1-\delta_{p_0}))$. 
To complete the UV-subtraction, we treat the 
$(1-\delta_{q_0})$ pieces of the first two terms as above:
in the second term we can once again use \eq\nr{piDec}
for $\sum\hspace*{-3.5mm}\int\hspace*{-0.5mm}_P\frac1{P^2(P-Q)^2}$,
while in the first term, we perform the analogous decomposition
\be \la{piTildeDec}
\sumint{P}\! \frac1{[P^2]^\e(P\!-\!Q)^2} =
\frac{\bar\beta}{[Q^2]^{2\e\!-\!1}}
+\frac{I_1}{[Q^2]^\e}
+\Delta\tilde\Pi(Q)
\ee
where we have again identified the UV-leading 4d 1-loop propagator
$\bar\beta=G(\frac{3-d}2,1,d+1)$ and the sub-leading behavior containing
the 1-loop tadpole sum-integral coming from only one routing of the
large external momentum, such that the UV-subtracted remainder
$\Delta\tilde\Pi(Q)\propto 1/Q^2$ in this case.

The three terms of \eq\nr{BN2} can hence immediately 
be rewritten (keeping the relative ordering of terms for clarity) 
as 4+4+3=11 terms as
\ba \la{BN3}
\lefteqn{B_N =}
\nn&&\hspace*{-6mm}
\beta \( A(N,\e,1)
+\bar\beta I_{N-1+2\e}
+I_1 I_{N+\e} \) 
+B_N^{IV}
\nn&&\hspace*{-6mm}
+2I_1 \( A(N,1,1)
+\beta I_{N+\e}
+2I_1I_{N+1}\)
+B_N^{III}
\nn&&\hspace*{-6mm}
+\sumint{PQ}\frac{\Delta\Pi(P)\delta_{p_0}\delta_{q_0}}{[Q^2]^N(P-Q)^2}
+B_N^{II}
+B_N^I\;,
\ea
where we have identified 1- and 2-loop vacuum sum-integrals 
$I_s$ and $A(s_1,s_2,s_3)$ for which explicit analytic results are
given in \se\ref{se:6},
and defined
\ba \la{bn1}
B_N^I &\equiv&
\sumint{P}\sumintp{Q}\frac{\Delta\Pi(P)}{[Q^2]^N(P-Q)^2}
\;,\\ \la{bn2}
B_N^{II} &\equiv&
\sumintp{P}\sumint{Q}\frac{\Delta\Pi(P)\delta_{q_0}}{[Q^2]^N(P-Q)^2}
\;,\\ \la{bn3}
B_N^{III} &\equiv& 2I_1\,
\sumintp{Q}\frac{\Delta\Pi(Q)}{[Q^2]^N}
\;,\\ \la{bn4}
B_N^{IV} &\equiv& \beta\,
\sumintp{Q}\frac{\Delta\tilde\Pi(Q)}{[Q^2]^N}\;,
\ea
where the primed sums denote $\sum_n^\prime=\sum_{n\neq 0}$.

The sum-integral that has been written out explicitly in \eq\nr{BN3}
can also be trivially solved by adding scale-free integrals
that vanish in dimensional reduction, viz
\ba
\lefteqn{\sumint{PQ}
\frac{\Delta\Pi(P)\,\delta_{p_0}\,\delta_{q_0}}{[Q^2]^N(P-Q)^2}}
\nn
&=&
\sumint{PQR}\frac{\delta_{p_0}\,\delta_{q_0}}{[Q^2]^N(P-Q)^2R^2(P-R)^2}
\nn
&=& 
T\,G(N,1,d)\,\sumint{PR}\frac{\delta_{p_0}}{[P^2]^{N+1-d/2}R^2(P-R)^2}
\nn
&=& 
T\,G(N,1,d)\,A(N+1-d/2,1,1) \;.
\ea

One of the four sum-integrals $B_N^{I-IV}$,
namely $B_N^{II}$,
containing a 3d 1-loop sub-integral where the zero-component of
the external momentum plays the role of a mass,
can be simplified using integration-by-parts (IBP) identities that
are so profitably employed in perturbative computations at $T=0$,
as witnessed by numerous contributions at this conference. 
This essentially automatizes the further IR subtractions that
would otherwise have been necessary for this term
since it still contains the zero-mode of its $1/[Q^2]^N$ propagator, 
see \cite{phi4}.
In fact, the IBP-reduction terminates with two master integrals,
(one of which is the massive 1loop tadpole G(1,d) given in \se\ref{se:6}),
\ba
\lefteqn{ \sumint{Q}\frac{\delta_{q_0}}{[Q^2]^N (P-Q)^2} }
\nn&&\hspace*{-5.5mm}=
T\intt{\vq}
\frac1{[\vq^2+p_0^2][(\vp-\vq)^2]^N}
\nn&&\hspace*{-6mm}\stackrel{\mbox{\tiny IBP}}{=} 
\frac{b_N(d,p_0^2/P^2)}{[P^2]^{N-1}}
\,T\,\intt{\vq}
\frac1{[\vq^2+p_0^2](\vp-\vq)^2}
\nn&&\hspace*{-6mm}\hphantom{\stackrel{\mbox{\tiny IBP}}{=}}
+\frac{a_N(d,p_0^2/P^2)}{[P^2]^N}  
\,T\,\intt{\vq}
\frac1{\vq^2+p_0^2}
\nn&&\hspace*{-5.5mm}=
\frac{b_N(d,p_0^2/P^2)}{[P^2]^{N-1}} 
\,\sumint{Q}\frac{\delta_{q_0}}{Q^2 (P-Q)^2}
\nn&&\hspace*{-5.5mm}\hphantom{=}
+\frac{a_N(d,p_0^2/P^2)}{[P^2]^N}\,|p_0|^{d-2}\,T\,G(1,d)\;.
\ea
The values $b_1(d,x)=1$ and $a_1(d,x)=0$ follow by definition, 
while utilizing e.g. FIRE \cite{fire}, one obtains
the polynomials
\ba
b_2(d,x)&=&(d-3)(2x-1) 
\nn
a_2(d,x)&=&d-2
\nn
b_3(d,x)&=&\frac12(d-3)(4(d-5)x(x-1)+d-4)
\nn
a_3(d,x)&=&\frac12(d-2)(d-5)(2x-1)
\ea
that we need below -- profiting from the observation
that $b_{N>1}\propto (d-3)$, since we will only be interested 
in constant terms at $d=3$.
We will also make use later of the generic structure
\be \la{aNn}
a_N(3,x)=\sum_{n=0}^{N-2}a_{N,n}\,x^n
\ee
with, in particular, $a_{20}=a_{30}=1$ and $a_{31}=-2$. 

\section{Treatment of finite terms at $d=3$}
\la{se:4}

After the subtraction procedure as outlined above, 
all that is left to do is to perform the four remaining
sum-integrals in \eqs\nr{bn1}-\nr{bn4}, which are defined such that
they are convergent and hence only need to be evaluated
in $d=3$, dropping terms of ${\cal O}(\e)$. 
This will enable us to perform the discrete sums explicitly.
Furthermore, as already mentioned in \se\ref{se:2},
it is now profitable to transform to coordinate
space, as there will be fewer integrations.

Writing the denominators in terms of their
(spatial) Fourier transforms with the help of 
\eqs\nr{ft1} and \nr{ft3}, the integration over spatial momenta
is trivial.
Using the values at $d=3$ of $\beta=\Gamma(\e)/(4\pi)^2$, 
$\bar\beta=-1/(64\pi^2)$,
$I_1=T^2/12$ and $G(1,3)=-1/(4\pi)$ from \se\ref{se:6},
we obtain the one-dimensional integral representations
\ba
B_N^I|_{\e=0} \!\!&=&\!\!
\frac{T^3\,2^{1\!-\!N}}{\Gamma(N)(4\pi)^3}
\int_0^\infty\!\!\!\!dr\,r^{N\!-\!3}\sum_{p_0}\Delta\pi(\bar r,\bar p_0)
\nn&\times&\!\!
\sum_{q_0\neq0} e^{-(|p_0|+|q_0|+|q_0-p_0|)r}
\frac{f_N(|q_0|r)}{|q_0|^{N-1}}
\,,\nn
B_N^{II}|_{\e=0} \!\!&=&\!\! 
-\frac{2T^3}{(4\pi)^3}\,\sum_{n=0}^{N-2}
\int_0^\infty\!\!\!\!dr\,
\frac{r^{N+n-2}\,a_{N,n}}{2^{N+n}\Gamma(N+n)}
\nn&\times&\!\!
\sum_{p_0\neq0}\frac{e^{-2|p_0|r}f_{N+n}(|p_0|r)\Delta\pi(\bar r,\bar p_0)}
{|p_0|^{N-n-2}}
\,,\nn
B_N^{III}|_{\e=0} \!\!&=&\!\! 
\frac{T^4\,2^{1-N}}{6\Gamma(N)(4\pi)^2}
\int_0^\infty\!\!\!\!dr\,r^{N-2}
\nn&\times&\!\!
\sum_{p_0\neq0}\frac{e^{-2|p_0|r}f_N(|p_0|r)\Delta\pi(\bar r,\bar p_0)}
{|p_0|^{N-1}}
\;,\nn
B_N^{IV}|_{\e=0} \!\!&=&\!\! 
\frac{T^2\,2^{1-N}}{\Gamma(N)(4\pi)^4}
\int_0^\infty\!\!\!\!dr\,r^{N-4}
\nn&\times&\!\!
\sum_{p_0\neq0}\frac{e^{-2|p_0|r}f_N(|p_0|r)
\Delta\tilde\pi(\bar r,\bar p_0)}
{|p_0|^{N-1}}
\;,\nonumber
\ea
where the coefficients $a_{N,n}$ have been defined in \eq\nr{aNn},
we have started to use dimensionless variables 
$\bar r\equiv 2\pi Tr$, $\bar p_0\equiv p_0/(2\pi T)$ 
and used the abbreviation $\Delta\pi(\bar r,\bar p_0)$ for
\be \la{abbrev1}
\sum_{r_0} e^{-(|r_0|+|r_0-p_0|-|p_0|)r} 
-\(|\bar p_0|+\frac1{\bar r}\) -\frac{\bar r}3
\ee
as well as $\Delta\tilde\pi(\bar r,\bar p_0)$ for the combination
\ba \la{abbrev2}
2\sum_{r_0} e^{-(|r_0|+|r_0-p_0|-|p_0|)r}\,\(|r_0|r+1\) -
\mbox{\hspace{5mm}}\nn
-\frac1{\bar r}\(p_0^2r^2+3|p_0|r+3\) -\frac{\bar r}3\(|p_0|r+1\) \,.
\ea

Using \eqs\nr{Li}-\nr{genericSum}
it is now straightforward to evaluate all sums,
whence \eq\nr{abbrev1} becomes in fact independent of $|p_0|$,
and in $B_N^I$ it is actually simplest to first sum
over $p_0$.
This leaves us with one-dimensional integrals over 
products of polylogarithms, logarithms and hyperbolic
functions.

\section{Numerical evaluation for $N=2$ and $3$}
\la{se:5}

While it would be most desirable to obtain expressions
for $B_N^{I-IV}$ for general $N$, in practice we are
forced to evaluate them at fixed $N$. While a number
(but not all) of them could be evaluated in closed form,
it is perhaps simplest to treat them all on the same
basis, i.e. in a numerical approximation, for which
we simply use the built-in routines of Mathematica \cite{math}.

For the special case $N=2$ we get
\ba
B_2^{I} &=& \!\!\!
-\frac{T^2}{(4\pi)^4}\times 0.0269726622737(1)
+{\cal O}(\e)\,, 
\nn
B_2^{II} &=& \!\!\!
\hphantom{-}\frac{T^2}{(4\pi)^4}\times 0.0134942763002(1)
+{\cal O}(\e)\,,
\nn
B_2^{III} &=& \!\!\!
-\frac{T^2}{(4\pi)^4}\times 0.0042655281176(1)
+{\cal O}(\e)\,, 
\nn
B_2^{IV} &=& \!\!\!
-\frac{T^2}{(4\pi)^4}\times 0.0004627085472(1) 
+{\cal O}(\e)\,.
\nonumber
\ea

At $N=3$, the numerical values are
\ba
B_3^{I} &=& 
-\frac{1}{(4\pi)^6}\times 0.0512974438185(1) 
+{\cal O}(\e)\,,
\nn
B_3^{II} &=& 
-\frac{1}{(4\pi)^6}\times 0.0163807421945(1) 
+{\cal O}(\e)\,,
\nn
B_3^{III} &=& 
-\frac{1}{(4\pi)^6}\times 0.0114452205501(1)
+{\cal O}(\e)\,,
\nn
B_3^{IV} &=& 
-\frac{1}{(4\pi)^6}\times 0.0035204424540(1) 
+{\cal O}(\e)\,.
\nonumber
\ea

We now have all the ingredients at our disposal to obtain 
the results for the two special cases $B_2$ and $B_3$. 
Adding up all contributions according to \eq\nr{BN3}, the
final outcome is presented in the concluding section.

\section{Building blocks of the computation}
\la{se:6}

In order to not clutter the main line of derivation with well-known
expressions, let us here collect a few simple results that were 
used in the previous sections. These are mainly the formulae for
analytically known (sum-) integrals, as well as our definitions
for the spatial Fourier transforms.

We have used the following zero-temperature integrals:
the 1-loop massive tadpole
\be
G(s,d) \equiv \intt{\vq}\frac1{[\vq^2+1]^s}
= \frac{\Gamma(s-\frac{d}2)}{(4\pi)^{d/2}\Gamma(s)}\,,
\ee
the 1-loop massless propagator
\ba
G(s_1,s_2,d) \equiv \(p^2\)^{s_{12}-\frac{d}2}\intt{q}
\frac1{[q^2]^{s_1}[(q-p)^2]^{s_2}}\mbox{\hspace{10mm}}
\nn=
\frac{\Gamma(\frac{d}2-s_1)\Gamma(\frac{d}2-s_2)\Gamma(s_{12}-\frac{d}2)}
{(4\pi)^{d/2}\Gamma(s_1)\Gamma(s_2)\Gamma(d-s_{12})} \;,\mbox{\hspace{6.5mm}}
\nonumber\ea
and the 2-loop tadpole 
(see e.g. \cite{tad2})
\ba
N(s_1,s_2,s_3) \equiv  
\intt{\vp\vq}
\frac1{[\vp^2+1]^{s_1}[\vq^2+1]^{s_2}[(\vp-\vq)^2]^{s_3}}
\nn= 
\frac{\Gamma(s_{13}-\frac{d}2)\Gamma(s_{23}-\frac{d}2)
\Gamma(\frac{d}2-s_3)\Gamma(s_{123}-d)}
{(4\pi)^d\Gamma(s_1)\Gamma(s_2)\Gamma(d/2)\Gamma(s_{1233}-d)}
\nonumber
\ea
where $s_{abc...} \equiv s_c+s_b+s_c+...\;$.

When dealing with discrete sums,
Zeta functions and polylogarithms enter through 
\be \la{Li}
\zeta(s) \equiv \sum_{n=1}^\infty n^{-s} 
\;\;,\;\;
\mbox{Li}_s(x) \equiv \sum_{n=1}^\infty \frac{x^n}{n^s} \,
\ee
with $\mbox{Li}_1(x)=-\ln(1-x)$,
while hyperbolic functions appear via ($m\in\mathbb{Z}$)
\ba \la{coth}
\sum_{n=-\infty}^\infty e^{-(|n|+|n-m|-|m|)r}
= |m|+\coth(r) \mbox{\hspace{5mm}}
\\
\la{genericSum}
\sum_{n=-\infty}^\infty e^{-(|n|+|n-m|-|m|)r} p(|n|)
=\sum_{n=0}^{|m|}p(n)+
\mbox{\hspace{0mm}}\nn
{}+p(-\partial_{2r})\frac1{e^{2r}\!-\!1}
+e^{2|m|r}p(-\partial_{2r})\frac{e^{-2|m|r}}{e^{2r}\!-\!1}
\mbox{\hspace{0mm}}
\ea
where $p(x)$ can be a polynomial. 
Note that \eq\nr{coth} is
just a special case of \eq\nr{genericSum}.

We have used the following sum-integrals:
the 1-loop bosonic tadpole
\be \la{I}
I_s \equiv \sumint{Q} \frac1{[Q^2]^s} 
= \frac{2T\,\zeta(2s-d)}{(2\pi T)^{2s-d}}\,G(s,d)
\ee
and a specific 2-loop tadpole
\ba
A(s_1,s_2,s_3) \equiv \sumint{PQ} 
\frac{\delta_{q_0}}{[Q^2]^{s_1}[P^2]^{s_2}[(P-Q)^2]^{s_3}}
\nn=
\frac{2T^2\,\zeta(2s_{123}-2d)}{(2\pi T)^{2s_{123}-2d}}\,N(s_2,s_3,s_1)\,.
\ea

Finally, let us define the $d$\/-dimensional (inverse, spatial) 
Fourier transforms ${\cal F}_s$ as
\be \la{ft1}
\frac1{[\vq^2+q_0^2]^s}
\equiv \int d^d\vr\,e^{i\vq\,\vr}\,r^{2s-d}\,{\cal F}_s(\sqrt{q_0^2r^2},d)
\ee
where, using a unit vector $\vec{e}$,
\ba \la{ft2}
{\cal F}_s(m,d) \equiv
\intt{\vp}e^{-i\vp\,\vec{e}}\,\frac1{[\vp^2+m^2]^s} 
= \mbox{\hspace{15mm}}
\nn
= \frac1{(2\pi)^d}\,\frac{2\pi^{d/2}}{\Gamma(d/2)}
\int_0^\infty\!\!\!\!dp\,\frac{p^{d-2}\sin(p)}{[p^2+m^2]^s} \;.
\ea
At $d=3$, ${\cal F}_s$ reduces to a modified Bessel function of second kind:
\ba \la{ft3}
{\cal F}_s(m,3) = \frac{e^{-m}}{4\pi(2m)^{s-1}\Gamma(s)}\,f_s(m)
\nn
f_s(m) \equiv \sqrt{\frac{2m}{\pi}}\,e^m\,K_{3/2-s}(m) \;.
\ea
The particular cases that we need read
$f_1(m)=f_2(m)=1$,
$f_0(m)=f_3(m)=(1+1/m)$ and
$f_{-1}(m)=f_4(m)=(1+3/m+3/m^2)$.

\section{Conclusions}
\la{se:7}

\newcommand{\gammaE}{\gamma_{\small\rm E}}
Collecting all terms 
we finally re-derive
\ba
B_2 &=& \frac{T^2\,(4\pi T^2)^{-3\e}}{8(4\pi)^4\e^2}
\lk 1+b_{21}\e+b_{22}\e^2+{\cal O}(\e^3)\rk
\nn
b_{21} &=& \frac{17}6+\gammaE+2Z_1\!'
\\
b_{22} &=&  \frac{131}{12} +\frac{31\pi^2}{36} +8\ln(2\pi)
-\frac{9\gammaE}{2} 
\nn &-&
 \frac{15\gammaE^2}{2}
 +(5+2\gammaE)Z_1\!'
 +2Z_1\!''-16\,\gamma_1
\nn &+&
 \frac{4\zeta(3)}9 -0.145652981107(4)
\ea
which coincides with the result presented in \cite{phi4}.
We also obtain the new result
\ba
B_3 &=& -\frac{(4\pi T^2)^{-3\e}}{12(4\pi)^6\e^2}
\lk 1+b_{31}\e+b_{32}\e^2+{\cal O}(\e^3)\rk
\nn
b_{31} &=& \frac92+3\gammaE-6\zeta(3)
\\
b_{32} &=& \frac{79}4 +\frac{13\pi^2}{12} -\frac{\pi^4}{90}
+\frac{27\gammaE}2 -\frac{27\gammaE^2}2 
\nn &+&
18\gammaE\zeta(3) -41\zeta(3)
-12\zeta(3)Z_1\!'-24\zeta'(3)
\nn&-&
36\gamma_1 -\frac{4\zeta(5)}3 +0.991726188205(4) \;.
\ea
We have abbreviated some derivatives of negative zeta values as
$Z_n\!'\equiv\frac{\zeta'(-n)}{\zeta(-n)}$,
$Z_n\!''\equiv\frac{\zeta''(-n)}{\zeta(-n)}$
and used the first of the Stieltjes constants defined
via $\zeta(s)=1/(s-1)+\sum_{n=0}^\infty(1-s)^n\gamma_n/n!$.

In closing, let us express our hope 
that a further generalization of the techniques 
presented here and possibly an automatization of the subtraction 
procedure will lead towards a computer-algebraic treatment
of multi-loop sum-integrals, in order to greatly streamline
higher-order perturbative computations in finite-temperature
field theories.

\section*{Acknowledgements}

This work was supported by the Deutsche Forschungsgemeinschaft
(DFG) under contract no.~SCHR 993/2-1.
The work of Y.S. is supported by the Heisenberg Programme
of the DFG.

\end{document}